\RequirePackage{fix-cm}
\documentclass[12pt]{article}
\usepackage{amsmath,amssymb,amsfonts}

\begin{document}
\title{Marder's Two-Fluid Dark Energy Cosmological Models In Saez-Ballester Theory of Gravitation}

\author{D.D.Pawar, Y.S.Solanke \\
School of Mathematical Science,Swami Ramanand Teerth Marathwada\\
 University, Nanded, 431606 (India), E-mail: dypawar@yahoo.com\\
 Department of Mathematics, Mungasaji Maharaj Mahavidyalaya,\\
Darwha,Dist. Yeotmal 445202, E-mail: yadaosolanke@gmail.com}
\date{Received: date / Accepted: date}
\maketitle
\begin{abstract}
The present paper deals with cylindrically symmetric metric in the form of Marder (1958) with  Saez-Ballester theory of gravitation in the presence of perfect fluid and dark energy. In order to obtain the deterministic solution of the field equations we have assumed that the expansion scalar in the model is proportional to the Eigen value of the shear tensor. We have also assumed that the two sources, here the perfect fluid and dark energy interact minimally with separate conservative parts of their energy momentum tensors together with the constant EoS parameter of the perfect fluid. The role of the dark energy in the present model with variable equation of state parameter is stuied more in detail. Some physical properties of model are also discussed.

\end{abstract}

\section{Introduction}
\label{intro}
The most remarkable astrophysical observations in the modern cosmology have  revolutionized our understanding about cosmology. According to the cosmologists our current universe is not only expanding but also accelerating. The direct evidence comes from distance measurements and analysis of type Ia supernovae (SN Ia), measurements of cosmic microwave background as well as large scale structure strongly suggest that present universe is dominated by the standard candle known as dark energy and it is due to because of cosmic accelerated expansion of the universe \cite{1}-\cite{9}. According to Einstein’s general theory of relativity in order to have such type of accelerated expansion of the universe, it is required to introduce new component to matter or the perfect fluid distribution of the universe with a large negative pressure. This new component most commonly known as dark energy. Thus from the recent observations obtained by the cosmologists it has been realized that without dark energy we can not explain the universe. The exact nature of the dark energy is known to be very homogeneous and not interact with any other fundamental forces except gravity. The form of the dark energy is not very dense so it is difficult to detect in the laboratory for the cosmologists. As the nature of dark energy and dark matter is unknown many radically different models have been proposed such as quintessence, tachyon, chaplygin gas,  as well as generalized chaplygin gas etc \cite{10}-\cite{15}. Having some limitations in Einstein theory of general relativity since it does not seems to resolve some of the important problems in the cosmology such as dark matter or the missing matter problems many researchers attracted towards the alternative theories of gravitation. Brans Dicke theory, Barber’s self creation theory, Saez-Ballester theory of gravitation are some of the alternative theories of gravitations \cite{16}-\cite{25}. Also,different two fluid models discussed by some researchers \cite{26}-\cite{29} \\
    The present paper deals with Saez-Ballester theory of gravitation. In Saez-Ballester theory of gravitation metric is coupled with a dimensionless scalar field $\phi$ in a simple manner. This $\phi$-coupling gives a satisfactory description of the weak fields. Inspite of the dimensionless character of the scalar field, an antigravity regime appears in this theory \cite{30}-\cite{35}. In order to obtain the deterministic solution we have assumed that two sources of the perfect fluid and dark energy interact minimally with independent conservation of their energy momentum tensors with constant EoS parameter of the perfect fluid.. We have investigated cylindrically symmetric cosmological model in the form of Marder (1958) . We have obtained some physical parameters and also discussed their physical behaviors \cite{36}-\cite{39}.

\section{Model and Field Equations}
We consider cylindrically symmetric metric in the form of  Marder (1958) given by
\begin{align}
ds^2=A_1^2(dx^2-dt^2)+A_2^2 dy^2+A_3^2 dz^2
\end{align}
  Where the metric potentials $A_1,A_2,A_3$  are the functions of cosmic time. Here it is important to note that by using the co-ordinate transformations $t\rightarrow \int A_1(t)dt$   metric given by equation (1) can turned into Bianchi type I. But for the sake of simplicity in the present paper we retain the metric given by equation (1).\\
The field equations of the Saez- Ballester scalar tensor theory are

 \begin{align}
 R_{ij}-\frac{1}{2}Rg_{ij}-\varpi \phi ^n\left(\phi_{,i}\phi_{,j}- \frac{1}{2}g_{ij}\phi_{,k}\phi^{,k}\right) =-T_{ij}
\end{align}
 Where the scalar field $\phi$   satisfying the equation
 \begin{align}
2\phi^n\phi_{,i}^{,i}+n\phi^{n-1}\phi_{,k}\phi^{,k}=0\\ and\quad
T_{ij}=T_{ij}^{(m)}+T_{ij}^{(de)}
\end{align}
is the overall energy momentum tensors with $T_{ij}^{(m)}$  as the energy momentum tensor of the ordinary matter or the perfect fluid and $ T_{ij}^{(de)}$ as the energy momentum tensor of the dark energy component. These are respectively given by

\begin{align}
T^{(m)j}_i=[T^4_4,T^1_1,T^2_2,T^3_3]=diag[-\rho^{(m)},p^{(m)},p^{(m)},p^{(m)}]\\
=diag[-1,\omega^{(m)},\omega^{(m)},\omega^{(m)}]\rho^{(m)},\nonumber\\
T^{(de)j}_i=[T^4_4,T^1_1,T^2_2,T^3_3]=diag[-\rho^{(de)},p^{(de)},p^{(de)},p^{(de)}]\\
=diag[-1,\omega^{(de)},\omega^{(de)},\omega^{(de)}]\rho^{(de)}.\nonumber
\end{align}
Where $\rho^{(m)},p^{(m)}$   are the energy density and pressure of the perfect fluid component respt. While$\rho^{(de)},p^{(de)}$  are the energy density and pressure of the DE component respectively where as $\omega^{(m)}=\frac{p^{(m)}}{\rho^{(m)}}$  and $\omega^{(de)}=\frac{p^{(de)}}{\rho^{(de)}}$  . Also in equation (2)$\varpi$   and n  are constants.\\
Now, the Saez-Ballester field equations (2) and (3) for the metric (1) with the help of equations (5) and (6) yield the following system of equations

\begin{align}
\frac{1}{(A_1)^2}\left(\frac{\ddot{A_2}}{A_2}+\frac{\ddot{A_3}}{A_3}-\frac{\dot{A_1}\dot{A_2}}
{A_1 A_2}+\frac{\dot{A_2}\dot{A_3}}{A_2 A_3}-\frac{\dot{A_1}\dot{A_3}}{A_1 A_3}\right)
-\frac{\varpi \phi^n \dot{\phi}^2}{2(A_1)^2}\\
=-\omega^{(m)}\rho^{(m)}-\omega^{(de)} \rho^{(de)},\nonumber\\
\frac{1}{(A_1)^2}\left(\frac{\ddot{A_1}}{A_1}+\frac{\ddot{A_3}}{A_3}-\frac{(\dot{A_1})^2}
{(A_1)^2} \right)-\frac{\varpi \phi^n \dot{\phi}^2}{2(A_1)^2}=-\omega^{(m)}\rho^{(m)}-\omega^{(de)} \rho^{(de)},\\
\frac{1}{(A_1)^2}\left(\frac{\ddot{A_1}}{A_1}+\frac{\ddot{A_2}}{A_2}-\frac{(\dot{A_1})^2}
{(A_1)^2} \right)-\frac{\varpi \phi^n \dot{\phi}^2}{2(A_1)^2}=-\omega^{(m)}\rho^{(m)}-\omega^{(de)} \rho^{(de)},\\
\frac{1}{(A_1)^2}\left(\frac{\dot{A_1}\dot{A_2}}
{A_1 A_2}+\frac{\dot{A_2}\dot{A_3}}{A_2 A_3}+\frac{\dot{A_1}\dot{A_3}}{A_1 A_3}\right)
+\frac{\varpi \phi^n \dot{\phi}^2}{2(A_1)^2}
=\rho^{(m)}+ \rho^{(de)},\\
\ddot{\phi}+\left(\frac{\dot{A_2}}{A_2}+\frac{\dot{A_3}}{A_3}\right)\dot{\phi}+\frac{n\dot{\phi^2}}{2\phi}=0.
\end{align}
   Also the energy conservation equation$T_{;j}^{ij}=0$   yields

\begin{equation}
\dot{\rho}^{(m)}+3[1+\omega^{(m)}]H\rho^{(m)}+\dot{\rho}^{(de)}+3[1+\omega^{(de)}]H\rho^{(de)}=0,
\end{equation}
\section{Solution of the field equations}
     The field equations (7) to (11) is a system of five independent equations in eight unknowns $A_1,A_2,A_3, \phi, \omega^{(m)},\rho^{(m)},\omega^{(de)}$\\and $\rho^{(de)}$. Therefore in order to obtain an explicit solution of the system we require three more suitable assumptions relating these three unknowns.
  Let us first assume the condition that the expansion scalar in the model is proportional to the shear scalar which leads to
\begin{equation}
A_{1}=(A_{2}A_{3})^{m}, m\neq1
\end{equation}\\
Differentiating equation (13) and with some little manipulation we have
\begin{equation}
\frac{\dot{A_1}}{A_1}=m\left(\frac{\dot{A_2}}{A_2}+\frac{\dot{A_3}}{A_3}\right)
\end{equation}\\
and\\
\begin{equation}
\frac{\ddot{A_1}}{A_1}=m\left[\frac{\ddot{A_2}}{A_2}+\frac{\ddot{A_3}}{A_3}-\frac{\dot{A_2}^2}{A_2^{2}}-\frac{\dot{A_3}^2}{A_3^2}\right]+m^2\left(\frac{\dot{A_2}}{A_2}+\frac{\dot{A_3}}{A_3}\right)^2
\end{equation}
Now comparing equation (8) and (9) we get
\begin{equation}
\frac{\ddot{A_2}}{A_2}-\frac{\ddot{A_3}}{A_3}=0.
\end{equation}
Substracting equation (9) from equation (7) we get
\begin{equation}
\frac{\ddot{A_2}}{A_2}-\frac{\ddot{A_1}}{A_1}+\frac{(\dot{A_1})^2}
{(A_1)^2}-\frac{\dot{A_1}\dot{A_2}}{A_1 A_2}+\frac{\dot{A_2}\dot{A_3}}{A_2 A_3}-\frac{\dot{A_1}\dot{A_3}}{A_1 A_3}=0.
\end{equation}
Equation (17) with the equations (14) and (15) gives
\begin{align}
\frac{\ddot{A_2}}{A_2}+\frac{\dot{A_2}\dot{A_3}}{A_2 A_3}=0.
\end{align}
For the sake of simplicity by setting the relation we have used  the substitutions
\begin{equation}
A_2 A_3=\lambda \quad and \quad \frac{A_2}{A_3}=\gamma.
\end{equation}
So that
\begin{equation}
(A_2)^2=\lambda \gamma \quad and \quad (A_3)^2=\frac{\lambda}{\gamma}.
\end{equation}
Then the equations (16) and (18)  respectively takes the form
\begin{equation}
\frac{d}{dt}(\frac{\lambda\dot{\gamma}}{\gamma})=0
\end{equation}
and
\begin{equation}
\frac{d}{dt}\left[\left(\frac{\dot{\lambda}}{\lambda}+\frac{\dot{\gamma}}{\gamma}\right)\lambda\right]=0
\end{equation}
Integrating equations (21) ,  (22)  we get
\begin{align}
\lambda=c_2 t+c_3 \quad i.e. \quad  A_2 A_3=c_2 t+c_3.
\end{align}
Where $c_2$  and $c_3$  are the constants of integrations.\\
  Thus from equations (13) and (23) we get a metric potential
\begin{equation}
A_1=(c_2 t+c_3) ^m.
\end{equation}
Equation (21) using equation (23) we get
\begin{equation}
\gamma=c_4(c_2 t+c_3) ^b.
\end{equation}
Where b  and  $c_4$  are the constants.\\
From equation (20) with the equations (23) and (25) we get the remaining metric potentials
\begin{equation}
A_2=c(c_2 t+c_3) ^{(\frac{b+1}{2})} \quad and \quad A_3=D(c_2 t+c_3) ^{(\frac{1-b}{2})}
\end{equation}
Where b, c, D are the constants.\\
  Thus our required cosmological model for the metric (1) is given by
\begin{equation}
ds^2=(c_2 t+c_3) ^{2m}(dx^2-dt^2)+c^2(c_2 t+c_3) ^{(b+1)}dy^2+D^2(c_2 t+c_3) ^{(1-b)}dz^2.
\end{equation}
Scalar field  $\phi$   from equation (11) with the help of the equations (24), and (26)  for the model (27)  is given by
\begin{equation}
\phi=\left[log(c_2 t+c_3) ^{k_0}\right]^\frac{2}{n+2}
\end{equation}
Where $k_0$ is constant.\\
To determine the energy density of the perfect fluid and DE components as well EoS parameters of the perfect fluid and DE components we have to assume following two more additional constraints.\\
As per the proposed assumption according to Akarsu and Kininc [11] let us suppose that the two sources of perfect fluid and dark energy interact minimally. Therefore energy conservation equation given by (12) can be split up into two separately additive conserved components which are given by
\begin{equation}
\dot{\rho}^{(m)}+3[1+\omega^{(m)}]H\rho^{(m)}=0,
\end{equation}
\begin{equation}
\dot{\rho}^{(de)}+3[1+\omega^{(de)}]H\rho^{(de)}=0.
\end{equation}
Finally we have assumed that the EoS parameter of the perfect fluid to be constant, i.e.
   $\omega^{(m)}=\frac{p^{(m)}}{\rho^{(m)}}= constant.$ \\
While $\omega^{(de)}$  is allowed to be a function of cosmic time since the current observational cosmological data from SN Ia, CMB and large scale structures mildly favor dynamically evolving dark energy crossing the phantom divide line (PDL).
\subsection{•Some physical parameter}
   The directional Hubble parameters of the model (27)  are defined as
\begin{equation}
H_x=\frac{\dot{A_1}}{A_1}=\frac{mc_2}{(c_2t+c_3)},H_y=\frac{\dot{A_2}}{A_2}=\frac{c_2(b+1)}{2(c_2t+c_3)},\nonumber\\
H_z=\frac{\dot{A_3}}{A_3}=\frac{c_2(1-b)}{2(c_2t+c_3)}.
\end{equation}
   Therefore mean Hubble parameter for the model is found to be
\begin{equation}
H=\frac{1}{3}(H_x+H_y+H_z)=\frac{(m+1)c_2}{3(c_2t+c_3)}
\end{equation}
The mean anisotropy parameter $\Delta$ of the expansion for the model is obtained as
\begin{equation}
\Delta=\frac{1}{3}\sum\left[\frac{H_i-H}{H}\right]^2=\frac{(2m-1)^2+3b^2}{2(m+1)^2}
\end{equation}
By the definition shear scalar and expansion scalar for the model (27) are respectively found to be
\begin{equation}
\sigma^2=\frac{(4m^2-4m+1+3b^2)c_2^2}{6(c_2t+c_3)^2}
\end{equation}
\begin{equation}
\theta=\frac{c_2(2m+1)}{c_2t+c_3}
\end{equation}
 Integraing equation (29) by using assumption of EoS parameter $\omega^{(m)}$  of the perfect fluid to be constant we get
\begin{equation}
\rho^m=\frac{k_2}{{(c_2t+c_3)}^{(m+1)}{[1+\omega^{(m)}]}}
\end{equation}
  Where $k_2$   being constant of integration.\\
  Equation (10) with the help of the equations (24), (26) and (35) gives the energy density of the DE component as
\begin{equation}
\rho^{(de)}=\frac{c_2^2[(n+2)^2(4m+1-b^2)+8\varpi{k}_0^2]}{4(n+2)^2(c_2t+c_3)^{2(m+1)}}
-\frac{k_2}{(c_2t+c_3)^{(m+1)}{[1+\omega^{(m)}]}}
\end{equation}
Equation (8) with the help of the equations (24), (26), (35) and (36) gives EoS parameter of the DE component as
\begin{equation}
\omega^{(de)}=-\frac{1}{\rho(de)}\left[\frac{c_2^2[(n+2)^2(b^2-4m-1)-8\varpi{k}_0^2]}{4(n+2)^2(c_2t+c_3)^{2(m+1)}}
+\frac{\omega^{(m)}k_2}{(c_2t+c_3)^{(m+1)}{[1+\omega^{(m)}]}}\right]
\end{equation}
\section{Discussion and conclusion:}
\quad The metric potentials $ A_1,A_2,A_3 $ all are finite at the initial moment but vanish when $ t=-\frac{c_3}{c_2} $    and increases with increase in cosmic time. Thus model have point type singularity at the initial epoch. Similarly the directional Hubble’s parameters as well as mean Hubble parameter are the functions of cosmic time t. All these parameters are finite at the early time of universe and diverge at initial singularity $ t=-\frac{c_3}{c_2} $ but vanish when cosmic time is infinite. The mean anisotropy parameter   of the model is constant throughout the evolution of the universe. The shear scalar   as well as expansion scalar   having the same behavior as that of  the Hubble parameters. The energy density of the perfect fluid is constant when cosmic time is zero and decreases with the expansion of the universe. In the present model EoS parameter of dark energy is a function of cosmic time. The nature of the dark energy depends on constants involved in the expression of $\omega^{(de)}$  . Also $\lim\frac{\sigma^2}{\theta^2}\neq0$ when $t\rightarrow\infty$.
In the present paper we have investigated cylindrically symmetric metric in the form of Marder by assuming that the two sources of the perfect fluid and dark energy interact minimally with EoS parameter of the perfect fluid to be constant. Also we have assumed the present paper is that current universe is dominated by the dark energy which can describe that the current universe is accelerating and consistent with observations.

\end{document}